\documentclass[11pt]{article}
\usepackage[latin9]{inputenc}
\usepackage{array}
\usepackage{multirow}
\usepackage{amsmath}
\usepackage{amssymb}
\usepackage{graphicx}
\usepackage[a4paper]{geometry}
\usepackage{setspace}
\makeatletter

\providecommand{\tabularnewline}{\\}

\numberwithin{equation}{section}
\numberwithin{figure}{section}
\newenvironment{lyxlist}[1]
	{\begin{list}{}
		{\settowidth{\labelwidth}{#1}
		 \setlength{\leftmargin}{\labelwidth}
		 \addtolength{\leftmargin}{\labelsep}
		 }}
	{\end{list}}

\usepackage{paralist}
 
\input{tcilatex}
\counterwithout{figure}{section}
\renewcommand{\thesection}{\arabic{section}.}
\usepackage{titlesec}
\titleformat{\section}
  {\normalfont\bfseries\centering}{\thesection}{0.3em}{}

\makeatother

\begin{document}
\title{\textbf{Consciousness, Quantum Mechanics, and the Limits of Scientific
Objectivism}\vspace{-0.1cm}
}
\author{{\Large John B. DeBrota \& Christian List\thanks{The paper presents the authors' shared programmatic vision; CL took
the lead in drafting the manuscript. The consciousness part of the
analysis draws on a talk given by CL on ``Consciousness and Objectivity''
at a workshop on ``Invariance and Objectivity'' at the Munich Center
for Mathematical Philosophy, October 2024, and a recent informal article
(List 2025a). The quantum mechanics part extends the authors' joint
work (DeBrota and List 2026). We are grateful to Philip Pettit for
discussion.}\smallskip{}
}\\
{\normalsize{} LMU Munich}{\Large\vspace{0.1cm}
}\\
{\footnotesize 14 April 2026}}
\date{{\small\vspace{-1cm}
}}

\maketitle
\noindent\noindent\rule{\linewidth}{0.4pt}

\noindent Consciousness and quantum mechanics are among the most puzzling
phenomena studied in the sciences. Some scholars suggest they are
related, though others think this claim commits a ``minimization
of mystery'' fallacy. The aim of this programmatic paper is to draw
attention to a less widely discussed parallel between consciousness
and quantum mechanics: both challenge the classical objectivist worldview
of science. Under certain \linebreak{}
assumptions, they are each in tension with a package of metaphysical
theses \textendash{} ``non-relationalism'', ``non-fragmentation'',
and ``one world'' \textendash{} that jointly make up that worldview.
This points to three distinct non-objectivist responses: the ``relationalist'',
``fragmentalist'', and ``many-subjective-worlds'' ones. We will
map out their pros and cons.{\small\vspace{-0.2cm}
}{\small\par}

\noindent\noindent\rule{\linewidth}{0.4pt}

\section{Introduction}

Consciousness and quantum mechanics are both thought to raise significant
challenges for science. Consciousness has been described as being
``at once the most familiar thing in the world and the most mysterious''
(Chalmers 1996, p. 3). Despite progress in neuroscience, we still
lack an accepted explanation of how it fits into the physical world.
Quantum mechanics makes some of the best-confirmed predictions we
find anywhere in science, for instance predicting with extraordinary
accuracy how electrons behave in magnetic fields, and it supports
technologies from lasers to MRI scanners, and yet there is nothing
even remotely resembling a consensus concerning its interpretation.
Both fields are riddled with paradoxes and unresolved problems: the
``hard problem of consciousness'' and the mind-body problem more
broadly, the puzzle of Schrödinger's cat, the Einstein-Podolsky-Rosen
paradox, and various ``no-go'' results, to name just a few. In fact,
some scholars have suggested that consciousness and quantum mechanics
are connected (see Gao 2022). According to Penrose (1994), for example,
consciousness cannot result from ordinary algorithmic processes but
may depend on non-computable processes, possibly involving resources
from quantum mechanics. Hameroff and Penrose's proposal is that consciousness
stems from quantum effects in tiny microtubules in the brain (2014).
Chalmers, on the other hand, attributes the tendency to connect these
two fields to the temptation to invoke a ``Law of Minimization of
Mystery'': ``consciousness is mysterious and quantum mechanics is
mysterious, so maybe the two mysteries have a common source'' (Chalmers
1995, section 5). But even Chalmers acknowledges possible connections.
Jointly with McQueen, he has recently revived the idea that the collapse
of the wave function is triggered by consciousness, an idea that goes
back to John von Neumann and Eugene Wigner but which has long been
dismissed (Chalmers and McQueen 2022). He further notes that the temptation
to see connections between the two fields ``is magnified by the fact
that the problems in quantum mechanics seem to be deeply tied to the
notion of observership, crucially involving the relation between a
subject's experience and the rest of the world'' (Chalmers 1996,
p. 333). 

In this paper, we want to draw attention to a less widely discussed
parallel between consciousness and quantum mechanics: both put pressure
on the classical ``objectivist'' worldview that is commonly associated
with science since the Enlightenment, and they do so in surprisingly
similar ways. Our claim is this:

\medskip{}

\noindent\textbf{A challenge for objectivism:} Consciousness and
quantum mechanics, at least under certain assumptions, are each in
tension with a package of metaphysical theses \textendash{} ``non-relationalism'',
``non-fragmentation'', and ``one world'' \textendash{} that jointly
make up the objectivist worldview of mainstream science.

\medskip{}

\noindent We will consider the argument for this claim in each of
the two domains and look at how we may need to revise the objectivist
worldview in response. Specifically, we will provide a roadmap of
three distinct kinds of non-objectivist approaches: ``relationalist'',
``fragmentalist'', and ``many-subjective-worlds'' approaches. 

Our analysis brings together insights from some recent ``no-go''
results concerning each domain: a ``quadrilemma for theories of consciousness''
(List 2025) and a ``heptalemma for quantum mechanics'' (DeBrota
and List 2026). These build on earlier results in the relevant fields,
especially Fine's (2005) analysis of first-personal facts and Bell's
theorem in quantum mechanics (Bell 1964), and so our analysis is indebted
to those earlier results too and to the bodies of work they have inspired.
Importantly, our focus will be on the big picture and on programmatic
issues raised by the challenge for objectivism. As it is often said,
the devil is in the detail, and there will be many details to which
we cannot do justice here. 

Scholars in the phenomenological tradition have also emphasized the
limitations of the objectivist worldview and have discussed problems
like the ones discussed here (for recent works, see, e.g., French
2023 and the volumes edited by Berghofer and Wiltsche 2020, 2023).
If we arrive in a similar place coming from a different angle, then
what we are arguing should be seen not so much as an attempt to reinvent
the wheel but rather as independent corroboration of broadly similar
ideas. On the other hand, the paper should also be of interest to
those who wish to resist the non-objectivist conclusion, because it
may help to clarify which premises one must reject in order to do
so.

To avoid any misinterpretation, we should stress that we are not suggesting
that consciousness is the key to solving the problems of quantum mechanics,
or that quantum mechanics is the key to solving the problems of consciousness.
What we are suggesting is something more modest: there is a \emph{structural}
parallel between the problems in the two areas, and making sense of
each set of problems may require a particular kind of departure from
the classical ``objectivist'' worldview of science. If there is
indeed such a parallel, this might lend support to a certain kind
of non-objectivist approach to science. It is this parallel and any
resulting lessons that we want to explore.

\section{The hard problem of consciousness}

Philosophers have long grappled with what David Chalmers has called
the ``hard problem of consciousness'' (1995, 1996).\footnote{Our exposition of the challenge draws on some of Christian List's
previous works, including List (2023, 2025) and especially List (2025a),
from which we have borrowed some wording.} While science is good at explaining many natural phenomena, from
the trajectories of the planets to the functioning of biological organisms,
science has struggled to explain how consciousness fits into the world.
The problem is not so much to explain why the capacity for cognition
and information processing has evolved in animals and humans. That
seems perfectly explicable. It will have been evolutionarily advantageous
for our ancestors to have that capacity: cognition and information
processing are likely to increase an organism's chances of survival
and procreation in a complex environment. The problem is to explain
why this is accompanied by subjective experiences, the sorts of experiences
that we each have from a first-person perspective. 

There is something it is like to be a conscious subject, for that
subject, as Nagel (1974) famously puts it. Why is this so? Why do
we each have a subjective perspective that we consciously experience,
over and above the apparently objective physical activity in our brains?
Why don't cognition and information processing take place ``in the
dark'', just as a computer engages in information processing without
presumably experiencing anything? It is generally thought that the
survival value of cognition and information processing can be explained
by reference to the \emph{objective} manifestations of the relevant
processes and does not depend on the subjective experiences accompanying
them: what it feels like to undergo those processes. 

The much-discussed thought experiment of a philosophical ``zombie''
helps to illustrate these points (Chalmers 1996, Kirk 2023). A philosophical
``zombie'' is a hypothetical creature that is objectively indistinguishable
from each of us but that subjectively experiences nothing. It outwardly
behaves in the same way each of us does. It produces the same verbal
responses we do. Its brain activity, according to EEG and fMRI scans,
is the same as ours. However, in this thought experiment, all of this
physical activity takes place without an inner stream of first-personal
consciousness. Such a hypothetical zombie would produce the same objective
responses to its environment as any one of us does, and thus it would
survive just as well. If, as Chalmers and others have argued, this
thought experiment is logically coherent (notwithstanding its far-fetchedness),
it helps us see that the first-person experiences that we each have
\textendash{} i.e., what it \emph{subjectively feels like} to undergo
the relevant cognitive processes \textendash{} is an unexplained ``add-on'':
something that cannot be explained just by pointing to the objective
manifestations of cognition and information processing. Expressed
slightly differently, conscious experience is first-personal and subjective,
while ordinary science is third-personal and objective. It seems that
a third-personal and objective explanation, of the sort science normally
offers, cannot explain a first-personal and subjective \emph{explanandum}.

One response to this problem is to argue that, to accommodate consciousness
in a philosophical or scientific worldview, we must assume that consciousness
is somehow fundamental rather than derived. Chalmers (1995, 1996)
has prominently defended one version of this response. Incorporating
consciousness into a scientific worldview, he suggests, would require
a theoretical innovation similar to the one made by James Clerk Maxwell
and others when they first postulated the existence of electromagnetic
fields. What Maxwell noted was that the inventory of properties that
exist in the physical world was richer than recognized in previous
physical theories such Newton's classical mechanics. There are not
only the Newtonian properties related to gravity and motion, but also
electromagnetic properties that were missing from Newton's theory.
Similarly, Chalmers claims, the world contains not only the generally
recognized physical properties, but also distinct \textquotedblleft phenomenal\textquotedblright{}
properties, which are responsible for conscious experiences. According
to Chalmers, it is the existence of those phenomenal properties that
accounts for the difference between the actual world, in which there
is consciousness, and the \textquotedblleft zombie world'' of the
thought experiment, in which everything is physically the same but
consciousness is absent. In effect, Chalmers proposes a modern, scientific
version of dualism, thereby updating the sort of traditional dualist
view associated with Descartes.

Chalmers's work has reframed the debate about consciousness around
the question of what kinds of properties must exist in the world to
account for consciousness along with everything else. The basic philosophical
disagreement, in this framing, is that between physicalists and non-physicalists.
Physicalists think that it is unnecessary to postulate any non-physical
properties over and above the ordinary physical ones: once all physical
properties are in place as they are, consciousness is an automatic
byproduct, albeit a highly complex one. Non-physicalists, like Chalmers,
object that this worldview fails to do justice to the intuition that
the thought experiment of a philosophical zombie is coherent (albeit
purely hypothetical) and that there remains a gap between what could
be explained by reference to physical properties alone and what must
be explained in order to account for consciousness. Consequently,
non-physicalists such as Chalmers think we must postulate phenomenal
properties as additional fundamental building blocks of the world
or give an enriched account of the nature of the physical.\footnote{A popular strategy is to postulate that there is one kind of fundamental
property, but that it has not only ordinary physical aspects \textendash{}
so-called functional or extrinsic ones \textendash{} but also phenomenal
or intrinsic aspects, which underpin consciousness. For an overview
of such ``monist'' theories, see Mørch (2024).}

These scholarly disagreements, however, obscure one important point
of agreement (as noted in List 2023): virtually all sides in the debate
share one key assumption, namely that we can think of reality as being
exhausted by a single unified and coherent world, which science and
philosophy seek to describe and explain and which is ``populated''
by all the facts to be explained. Call this the ``objectivist''
assumption; we will make it more precise later. It seems that almost
everyone in the analytic tradition, from the most hardline physicalist
at one end of the spectrum to the staunchest dualist at the other,
accepts this assumption or at least tacitly takes it for granted.
The main disagreements merely concern the precise ontological inventory
of that ``objective world'': what facts or properties populate it;
whether it is it exhausted by physical facts and properties alone;
or whether it contains something over and above the physical, such
as special phenomenal properties or properties that somehow possess
both intrinsic and extrinsic aspects. The heated nature of this debate
makes it easy to overlook the shared objectivist assumption that sits
in the background.

Why does this matter? It matters because all of the mainstream scientific
and philosophical theories of consciousness, including some of the
most popular non-physicalist theories such as Chalmers's, give us
a fundamentally third-personal or impersonal picture of the world.
They attempt, in effect, to describe and explain ``the world'' as
it would be seen by an omniscient, or highly informed, outside observer:
someone who is taking what Nagel (1986) has called \textquotedblleft the
view from nowhere\textquotedblright . 

The attempt to abstract away from any particular perspective is, of
course, a central feature of science as we know it. The goal is to
identify general patterns and regularities: ones that are invariant
under shifts in perspective, i.e., which remain stable as we move
from one perspective to another, and which thereby qualify as ``objective''.
Newton's laws of motion are examples. To identify such patterns and
regularities, we must abstract away from the particularities of any
specific objects, events, and perspectives. We must aim to arrive
at a non-perspectival worldview \textendash{} the sort of worldview
for which Nagel coined the term ``the view from nowhere''. The history
of science illustrates this process of abstraction, with the transition
from an anthropocentric to a geocentric to a heliocentric and ultimately
to a more universal view of the world, and by and large this has been
a success story.

Nevertheless, there appears to be at least one phenomenon that resists
this kind of objectivization or ``de-perspectivization'', and that
is the phenomenon of consciousness itself. Later we will consider
whether some quantum-mechanical phenomena fall into a similar category.
In the case of consciousness, the defining feature is subjective experience:
there is something it is like to be a conscious subject, for that
subject, as Nagel notes. This puts subjectivity at the heart of consciousness.
The phenomenologists in the tradition of Edmund Husserl (e.g., 1970)
have been well aware of this point (Zahavi 2025) and have very much
resisted the idea of a ``view from nowhere''. For them, the first-personal
character of experience is central and what we call ``the objective
facts'' are an emergent abstraction that corresponds broadly to intersubjective
agreement. The core of my, say Christian's, consciousness lies in
the fact that I find myself in a world in which there are first-personal
facts. I am conscious, I have certain experiences, I am in a particular
perceptual state, and so on. First-personal facts are irreducibly
subjective. They are \textquotedblleft centred\textquotedblright{}
around my perspective as an experiencing subject, and unlike objective
facts, they are not invariant under shifts in perspective. If we shift
from one perspective to another, the first-personal facts change.
The ``objective'' facts are broadly those that survive such shifts. 

If this is right, then the correct move to accommodate consciousness
in our worldview may not be to postulate a new class of properties
\textendash{} phenomenal properties \textendash{} within the objective
world as ordinarily understood, but rather to postulate irreducibly
first-personal facts as a feature of reality: facts that have an inbuilt
conscious perspective, facts which are ``centred'' around a subjective
perspective.\footnote{Nagel already recognized that just enriching the postulated inventory
of the objective world, such as along dualist lines, is insufficient.
In his words (quoted and discussed in Ratcliffe 2002, p. 356): ``The
broader issue between personal and impersonal, or subjective and objective,
arises also for a dualist theory of mind. The question of how one
can include in the objective world a mental substance having subjective
properties is as acute as the question how a physical substance can
have subjective properties.''} On this picture, ``{[}r{]}eality is not exhausted by the \textquoteleft objective\textquoteright{}
or impersonal facts but also includes facts that reflect a first-person
point of view\textquotedblright , as Kit Fine puts it (2005, p. 311).
Fine calls this thesis ``first-personal realism'', albeit without
endorsing it. 

A simple example of a first-personal fact, which holds where I, Christian
\textendash{} currently the narrator \textendash{} stand, is the fact
that I am seeing a beige computer screen as I am writing this right
now. This must not be confused with the third-personal fact that Christian
is seeing a beige computer screen at a specific point in time. The
latter is a third-personalized version of the given first-personal
fact, as it continues to hold even if we step away from Christian's
particular first-person perspective. The third-personal fact is true
for John and everyone else as much as it is true for Christian. It
is true, from where you stand, that Christian is seeing a beige computer
screen at this particular time. However, assuming that you are not
looking at a beige computer screen, ``I am seeing a beige computer
screen right now'' is not a first-personal fact for you.

Postulating first-personal facts is still very heterodox in both science
and analytic philosophy. According to the mainstream view, the first-person/third-person
distinction cannot be drawn at the level of facts at all but is just
a linguistic or cognitive distinction. There are first-personal \emph{modes
of representation} of some facts, but the facts themselves are always
third-personal or impersonal; facts themselves never come with an
in-built perspective. According to the orthodox view, there is no
distinct first-personal fact such as ``I am seeing a beige computer
screen right now''. The sentence in quotation marks is just a first-personal
way of expressing the impersonal fact that Christian is seeing a beige
computer screen at this particular point in time. 

Some philosophers, however, think that our picture of reality is incomplete
unless we accept that there are genuine first-personal facts, and
not just their third-personalized counterparts. We miss out on an
important aspect of reality if we don't recognize that there are irreducibly
first-personal facts. For philosophers in the phenomenological tradition,
as already noted, the first-person perspective is central (see Zahavi
2025). Once we recognize that the first-person perspective is our
window to the world and that all experience is first-personally given,
it becomes natural to think that any enquiry must begin with first-person
facts and can arrive at third-person facts only subsequently, by abstracting
away from the particularities of any first-person perspective (for
instance, by asking which facts are invariant under changes in the
first-person perspective). This was approximately Husserl's picture.
French (2023, p. 186) characterizes ``the subjective stance'' as
one of ``the core features of phenomenology''. Among analytically
oriented philosophers Lynn Rudder Baker (2013) is one of the few who
have defended the existence of first-person facts. She argues that
because the scientific worldview \textquotedblleft takes the world
to be impersonal'', it cannot properly \textquotedblleft accommodate
me\textquotedblright , as a conscious subject (2013, xiii\textendash xiv).
To accommodate subjects and their perspectives, she suggests, we must
accept that \textquotedblleft there are irreducible first-person facts''
(ibid.). Nagel, similarly, notes that a purely third-personal view
of the world leaves out ``the fact that I am Thomas Nagel\textquotedblright{}
(1965, p. 355; on parallels between Nagel and Husserl, see Ratcliffe
2002). Nagel further writes:
\begin{quote}
``The reductionist program that dominates current work in the philosophy
of mind is completely misguided, because it is based on the groundless
assumption that a particular conception of objective reality is exhaustive
of what there is.'' (Nagel 1986, p. 16) 
\end{quote}
Indeed, in a purely third-personal worldview, there is no room for
the fact that I am who I am, or that I occupy a particular perspective
on the world. There is only room for the fact that Christian occupies
a particular perspective, or that John occupies a particular perspective,
but as already noted, that's not the same as the fact that \emph{I}
occupy such-and-such perspective. 

More recently, however, there has been a growing interest in realism
about first-personal facts. Martin Lipman (2023), for example, asks:
``What shape must reality take for there to be a place for consciousness
in it?\textquotedblright{} (p. 530). In answer, he writes: ``for
the world to harbor consciousness, reality must include subjects that
are metaphysical standpoints and be such that some of the facts obtain
only from the standpoint of certain subjects and not others\textquotedblright{}
(pp. 530\textendash 531). He sums up the view by saying that ``reality
is constituted by subjective facts'' (p. 531). Others who have explored
similar views include \linebreak{}
Hellie (2013), Merlo (2016), Conitzer (2020), one of us (List 2023),
and Builes (2024). 

Yet, once we postulate first-personal facts, we face a significant
theoretical challenge. We must reconcile the existence of first-personal
facts with the objectivist view of reality that we traditionally find
in science and philosophy or alternatively revise that view. It is
this challenge that we now want to discuss. 

\section{The challenge for objectivism}

Let us start with the following thesis:

\medskip{}

\noindent\textbf{First-personal realism:} Reality includes, for each
conscious subject, some first-personal facts.

\medskip{}

We want to examine how this relates to the standard objectivist worldview
of science and philosophy. That worldview, we propose, consists of
three theses about the nature of reality, which we call ``one world'',
``non-fragmentation'', and ``non-relationalism'' (following the
terminology in List 2025 and DeBrota and List 2026). The three theses
are so widely accepted that they are seldom explicitly stated. The
first expresses the core idea that there is one objective world. 

\medskip{}

\noindent\textbf{One world: }Reality is exhausted by one world (which
we may think of as an objective world), not multiple worlds (which
we may think of as subjective ones).

\medskip{}

\noindent Although usually left implicit, this thesis is explicitly
discussed, for example, in Strawson (1966,\,part\,2,\,ch.\,3,\,section\,8),
who, in turn, refers to its Kantian origins. Commenting on those origins,
Moore (2016) quotes Kant's statement that ``all appearances lie,
and must lie, in \emph{one} nature'' (Kant, transl. by Smith 1929,
A216/B263) and interprets it as the thesis that ``the Cosmos constitutes
the whole of empirical reality'', where ``the Cosmos'' is defined
as ``the four-dimensional realm that we inhabit'' (Moore 2016, p.
934).\footnote{Moore points out that ``Kant himself allows for the possibility that
the Cosmos does not exhaust reality in a broader sense of \textquoteleft reality\textquoteright \,''(ibid.).
We will consider that possibility later in the present paper.}

The second thesis adds an important qualification about the nature
of any such world.

\medskip{}

\noindent\textbf{Non-fragmentation: }Any world (actual or possible)
is a coherent collection of facts. Equivalently, the total collection
of facts that obtain at any given world can be coherently coinstantiated. 

\medskip{}

\noindent This idea that any possible world must be coherent is very
much in the spirit of the Kantian picture mentioned above. The relevant
notion of a world is nicely articulated in the opening sentence of
Wittgenstein's famous \emph{Tractatus}: ``The world is everything
that is the case'', which he further clarifies by adding that ``{[}t{]}he
world is the totality of facts, not of things'' (1922, 1 and 1.1).
The coherence of any world, as required by non-fragmentation, is essential
to that notion. A world that is not a coherent totality of facts would
not be a possible world as conventionally understood, let alone a
candidate for being the actual world. Fine (2005) uses the name ``coherence''
for a version of the non-fragmentation thesis.

The third thesis adds another important qualification:

\medskip{}

\noindent\textbf{Non-relationalism:} Any fact within the collection
of facts that constitute reality is of the absolute form ``such and
such is the case'', not of the relative form ``such and such is
the case, relative to such and such''.

\medskip{}

This thesis can also be motivated by reference to Wittgenstein's understanding
of facts. Recall that, for Wittgenstein, a fact is \emph{something
that is the case}. It is something that is the case \emph{simpliciter},
not something that is the case only relative to something else. Examples
of facts are the following: there is life on Earth; carbon dioxide
contains one carbon atom and two oxygen atoms; London is a city; $1+1=2$.
Each of these facts holds \emph{simpliciter}, not relative to something
else. There is life on Earth, full stop. This is not merely the case
relative to something else. Carbon dioxide has the stated composition,
full stop, not just relative to something else. To understand the
difference between a fact in this absolute sense and a merely relative
fact, contrast ``carbon dioxide contains two oxygen atoms'' with
``carbon dioxide contains more oxygen atoms''. The first expresses
a fact in the absolute sense. The second invites the question: relative
to what? Relative to carbon monoxide, carbon dioxide contains more
oxygen atoms. But relative to carbon trioxide or sulfur trioxide,
it does not. Without any relativization parameter, the sentence ``carbon
dioxide contains more oxygen atoms'' does not express a fact in the
ordinary sense. Non-relationalism asserts that the facts making up
reality are of the absolute form ``such and such is the case'',
not of the relative form ``such and such is the case relative to
such and such''. Fine (2005) call this thesis ``absolutism''. 

Let us use the label ``objectivism'' for the conjunction of the
three theses just stated: one world, non-fragmentation, and non-relationalism.
Objectivism thus defined is an ontological rather than epistemological
thesis. That is to say, it is a thesis about what reality is like,
not a thesis about what we can know about reality and how we can do
so. If objectivism is true, then reality consists of one world, which
we can think of as a coherent and exhaustive collection of facts that
are instantiated at that world and where those facts have an absolute
rather than relative form. What we can come to know about reality,
and how we can do so, remains a separate question, which is epistemological
and not our concern here, although it is an important question in
its own right.

We now have all the background in place to see the tension between
first-personal realism and objectivism. The tension arises under the
reasonable assumption of non-solipsism:

\medskip{}

\noindent\textbf{Non-solipsism:} More than one conscious subject
is real; in particular, I am not the only conscious subject.

\medskip{}

\sloppy Our claim (originally defended in List 2025) is that first-personal
realism, non-solipsism, and objectivism are jointly inconsistent.\footnote{In contrast to the exposition in List (2025), we are here combining
three theses (one world, non-fragmentation, and non-relationalism)
into a single thesis (objectivism). Further, in List (2025), non-relationalism
was not stated as a separate thesis but was treated as a presupposition
of first-personal realism; see especially section III of that paper.
We will come back to this point later.} To see this, suppose we accept both first-personal realism and non-solipsism.
Then more than one conscious subject is real. For each conscious subject,
there are first-personal facts. Furthermore, given non-relationalism,
these first-personal facts obtain \emph{simpliciter}, not relative
to anything else. For example, the fact that I \textendash{} say Christian
\textendash{} am Christian is the fact that 
\begin{lyxlist}{(iii)}
\item [{(i)}] \noindent I am Christian, full stop. 
\end{lyxlist}
It is not merely the fact that 
\begin{lyxlist}{(iii)}
\item [{(ii)}] \noindent I am Christian, relative to Christian's perspective,
\end{lyxlist}
or the fact that
\begin{lyxlist}{(iii)}
\item [{(iii)}] relative to Christian's perspective, it is true to say
``I am Christian''.
\end{lyxlist}
Of course, what is expressed by sentences (ii) and (iii) is true,
but it is not a genuinely first-personal fact. Even from where John
and everyone else stands, it is true that ``relative to Christian's
perspective, it is true to say `I am Christian'{}''. Just as the
fact that Christian is seeing a beige computer screen at a specific
time is a third-personal fact, so the facts expressed with the help
of the ``relative to''-clauses, i.e., (ii) and (iii), are not genuinely
first-personal either. Only ``I am Christian'' expresses a first-personal
fact; ``I am Christian, relative to Christian's perspective'' or
``relative to Christian's perspective, it is true to say `I am Christian'{}''
do not.

With this qualification in place, it becomes evident that different
subjects' first-personal facts are mutually incompatible. As an illustration,
consider the first-personal fact, which holds from where I, Christian,
stand, that I am seeing a beige computer screen in front of me right
now. Suppose that you are not seeing a beige computer screen in front
of you right now; you are seeing something else. Then, from where
you stand, the following is true: ``I am not seeing a beige computer
screen in front of me right now''. The two first-personal facts (``I
am seeing a beige computer screen in front of me right now'', and
``I am not seeing a beige computer screen in front of me right now'')
cannot be co-instantiated as first-personal facts. Only their third-personalized
counterparts are compatible: Christian is seeing a beige computer
screen in front of him right now; Christian's interlocutor is not
seeing a beige computer screen in front of them right now. Or alternatively,
their relativized versions are compatible: ``relative to Christian's
perspective, it is true to say `I am seeing a beige computer screen
in front of me right now'{}'' and ``relative to the perspective
of Christian's interlocutor, it is true to say `I am not seeing a
beige computer screen in front of me right now'{}''. But these are
not first-personal facts; they are third-personalized or merely relational
counterparts of the given first-personal facts. So, we have established
the following:

\medskip{}

\noindent\textbf{Lemma:} Given non-relationalism, the first-personal
facts of different conscious subjects are not all mutually compatible.

\medskip{}

Since first-personal realism commits us to realism about first-personal
facts, not merely to realism about their third-personalized or relational
counterparts (noting again that we have assumed non-relationalism),
we get the conflict with the claim that reality is exhausted by one
coherent world. The total collection of facts constituting reality,
which includes different subjects' first-personal facts, is not coherent.
Therefore, we cannot simultaneously accept one world and non-fragmentation.
If reality is exhausted by a single world, this will be incoherent.
And if worlds are each supposed to be internally coherent, then reality
cannot be exhausted by a single such world.

Summarizing these observations, we get a no-go result, which is a
variant of the result in List (2025). A precursor of the result can
be found in Fine (2005).\footnote{Fine notes that first-personal realism conflicts with three other
theses that are similar to the ones discussed here: \textquotedblleft first-personal
neutrality\textquotedblright , which roughly corresponds to non-solipsism;
\textquotedblleft absolutism\textquotedblright , which corresponds
to non-relationalism; and \textquotedblleft coherence\textquotedblright ,
which corresponds to non-fragmentation or perhaps the conjunction
of non-fragmentation and one world. Fine does not explicitly state
the thesis of one world but seems to presuppose that reality is exhausted
by a single world.}

\medskip{}

\noindent\textbf{No-go result for consciousness:} First-personal
realism, non-solipsism, and objectivism are jointly inconsistent.
Any two of the three theses are jointly consistent.

\medskip{}

To see that any two of the three theses are jointly consistent, it
suffices to consider the possibilities of satisfying any pair of them.
The least compelling possibility is arguably the denial of non-solipsism.
If we are prepared to suppose that reality admits only a single conscious
subject \textendash{} myself \textendash{} then the issue of how to
reconcile different subjects' incompatible first-personal facts no
longer arises. We can then postulate a single coherent world that
accommodates all of my (the only subject's) first-personal facts along
with all facts that we would conventionally classify as third-personal
or impersonal, in a way that that is formally consistent with how
we have defined objectivism. Caspar Hare (2007) has defended such
a solipsist view, which he calls ``egocentric presentism''. While
the view is consistent (and it's important to acknowledge it as a
possible position), however, this is unlikely to be a route many would
want to take. As Fine notes, it seems ``metaphysically preposterous
that, of all the people there are, I am somehow privileged \textendash{}
that my standpoint is the standpoint of reality and that no one else
can properly be regarded as a source of first-personal facts'' (2005,
p. 285). 

The denial of first-personal realism, by contrast, is very popular.
It can be found in most mainstream theories of consciousness in analytic
philosophy, most notably in physicalist theories but also in Chalmers's
dualistic alternative, as already discussed in the previous section.
Scientific theories that seek to identify consciousness with certain
neuronal properties or patterns of brain activity also fall into this
camp, for instance those that say ``consciousness \emph{is} synchronized
brain activity in a certain frequency range''. Those theories embrace
standard scientific objectivism, and they are non-solipsistic. For
reasons of consistency, they must then reject first-personal realism;
and indeed, they do. Interestingly, with the exception of so-called
illusionist theories (Frankish 2016), which explicitly regard consciousness
as an illusion, the mainstream theories still claim that they consider
consciousness to be real.\footnote{Illusionism raises the question of whether there can be an illusion
without a conscious subject for whom it is an illusion. Illusionists
must, in effect, reject the commonsense reading of this question.} Nevertheless, by rejecting first-personal realism they are committed
to a certain form of anti-realism about first-personal facts, whether
or not they admit this. Anyone who is persuaded by the centrality
of subjectivity to consciousness may therefore be inclined to conclude
that those theories fail to accommodate consciousness properly. This
leaves the denial of objectivism as a serious possibility. 

\section{Non-objectivist approaches to consciousness}

Given that objectivism is the conjunction of three theses, any one
of them can in principle be relaxed while upholding the other two.
Perhaps the least radical departure from objectivism, in some respects,
is to uphold one world and non-fragmentation and to argue that first-personal
facts are only relative rather than absolute facts. We would thus
relax non-relationalism, but this relaxation would be restricted to
first-personal facts. The mutual incompatibility of different subjects'
first-personal facts would then be avoided insofar as there is no
conflict between it being true relative to Christian's perspective
that ``I am seeing a beige computer screen right now'' and this
not being true relative to someone else's perspective, such as John's.
But while this is a coherent position, it is unclear whether it does
justice to the nature of first-personal facts. Given the proposed
relaxation of non-relationalism, first-personal facts would only be
facts of a relativized sort, such as ``relative to Christian's perspective,
it is true to say `I am having Christian's experiences'{}'', and
those relativized facts are arguably not genuinely first-personal.
Even the total collection of such relativized facts would leave out
the fact that \emph{I am having Christian's experiences rather than
anyone else's}. To illustrate, consider the collection of facts: 
\begin{itemize}
\item relative to Christian's perspective, it is true to say ``I am having
Christian's experiences'';
\item relative to John's perspective, it is true to say ``I am having John's
experiences'';
\item relative to Alice's perspective, it is true to say ``I am having
Alice's experiences'';
\item and so on.
\end{itemize}
This collection of facts leaves open which experiences \emph{I} have.
It does not settle the question of whether I am Christian, or John,
or Alice, or someone else. This is because the relativized facts are
still inventorized by an Olympian observer adopting a view from nowhere;
they do not pick out any perspective at all.\footnote{Another way of expressing the same point is to say that the relativized
facts, unlike genuinely first-personal facts, do not settle what Hellie
(2013) has called the ``vertiginous question'': why am I having
\emph{my} experiences rather than anyone else's? Now, perhaps Hellie's
question has no satisfactory answer. But at least we might want to
insist that it is a fact \emph{that} I am having my experiences rather
than anyone else's, even if this is a brute fact for which one can
give no further explanation. The relationalist approach would not
be able to recognize such a brute fact. According to it, there would
only be the relative fact that relative to being Christian, it's true
to say ``I am having Christian's experiences'', but that's not the
fact we are after if we take the vertiginous question seriously.} Reconstruing first-personal facts as relational facts would thus
amount to a denial of first-personal realism in the originally intended
sense. Moreover, if we went along the relationalist route, we would
still have to answer the question of what exactly the relativization
parameter is. Presumably, it would be a ``subject'' or a ``subjective
perspective'', but then we would need to explain how subjects or
subjective perspectives fit into the world. Interestingly, this is
where Fine (2005) sees the greatest difficulties with the relationalist
position. He writes:
\begin{quote}
``If we opt for the relativist position, then we must take each subjective
reality to be given relative to a metaphysical subject or self. But
reality itself contains no metaphysical self. We therefore arrive
at the conception of the pure metaphysical self \textendash{} one
that stands outside the world and yet is that by which the world (or
the subjective world) is given. One can see why philosophers might
have been attracted to such a position, given that they wished to
give proper recognition to the multiplicity of different subjective
viewpoints that, in themselves, were without a point of view. But
the position is barely intelligible; and the mystery of the pure metaphysical
self no longer arises once we opt for the fragmentalist {[}as distinct
from the relativist{]} position.'' (p. 314)
\end{quote}
Indeed, most first-personal realists in the small scholarly literature
on the subject do not take the relationalist route. Setting aside
solipsism, the most popular route appears to be the relaxation of
non-fragmentation, as already anticipated in Fine's quoted passage.
This would allow us to uphold the view that reality is exhausted by
one world and that all facts making up that world, including all first-personal
facts, are non-relational. We would have to concede, however, that
the world is incoherent. This seems metaphysically costly, given that
the notion of coherent worlds is so central to logic, philosophy,
and science. Yet, the position does do justice to the idea that I
am having my own experiences \emph{simpliciter}, and not just in some
relativized sense. And indeed, if a first-personal realist rejects
both solipsism and relationalism and wishes to uphold the standard
idea of a single world, then they must embrace a fragmentalist worldview.
This is also the conclusion to which Fine (2005) is led by his analysis,
and so is Lipman (2023). To make fragmentalism more palatable, one
might say that although reality as a whole is a globally incoherent
collection of facts, it has locally coherent subcollections: coherent
fragments. Merlo (2016, p. 324) characterizes the relevant metaphysical
picture along similar lines: ``one could take all points of view
to be on a par vis-a-vis truth simpliciter by treating them as different
\textquoteleft fragments\textquoteright{} of an overall incoherent
totality of facts\textquotedblright . 

A third route is to accept that there is not just a single third-personal
or objective world, but many different first-personally centered or
subjective worlds, corresponding to different conscious subjects.
Each of these subjective worlds can then still be internally coherent
and individually made up of non-relational facts, so that we can uphold
non-fragmentation and non-relationalism. This route breaks with the
traditionally assumed identity ``reality = the world''. According
to it, reality consists of a collection of subjective worlds, not
a single objective world. This would be a many-worlds theory of consciousness,
or a many-first-personally-centred-worlds theory (as sketched in List
2023). The resulting worldview is obviously very heterodox, and metaphysically
costly in other respects. Arguing for it would require a balancing
of considerations. In particular, if first-personal realism, non-solipsism,
non-relationalism, and non-fragmentation seem too implausible to abandon,
then the many-subjective-worlds route becomes a plausible contender.
It is worth noting that a few philosophers have previously considered
the possibility of giving up the assumption of one world. Strawson
recognizes the question ``{[}w{]}hy only \emph{one} unified objective
world'' as a genuine one and acknowledges some rival possibilities,
including ``a multiplicity of objective worlds'' (1966, p. 151).
Goodman (1978, 1984) develops the ``pluriworldist'' view that there
are multiple actual worlds (Declos 2019), and Lewis (1986) famously
defends the modal realist thesis that many possible worlds are real,
even though only one is actual. (Lewis's account, however, does not
refer to subjective or first-personally centred worlds, and so his
proposal is still quite distinct from what we are discussing here.)
More recent contributions that give up the idea of one objective world,
including in the context of the mind-body problem, are Vacariu's (2005,
2016) account of ``epistemologically different worlds'', Honderich's
(2014) account of ``subjective physical worlds'', and Gabriel's
(2015) argument that ``the world'', understood as a unified totality
of everything, does not exist. 

All three non-objectivist routes \textendash{} the relational route,
the fragmentalist route, and the many-subjective-worlds route \textendash{}
are likely to be outside the metaphysical comfort zone of any proponent
of the standard objectivist worldview, all the more so because that
worldview has served science so well ever since the Englightenment.
It is therefore unsurprising that the dominant approach in the analytic
philosophy of consciousness would rather reject first-personal realism
than entertain a departure from objectivism. And indeed, if consciousness
were the only phenomenon for which such a departure might be needed,
then the move away from objectivism might seem \emph{ad hoc} and too
costly. However, this is where it is instructive to consider the case
of quantum mechanics.

\section{The EPR paradox and Bell's theorem}

To introduce the puzzles of quantum mechanics, it is helpful to begin
with the Einstein-Podolsky-Rosen (EPR) paradox (1935). Suppose we
set up a device that emits two maximally entangled particles travelling
in opposite directions, as illustrated in Figure 1.
\begin{figure}[h]
\caption{The setup in the EPR paradox}

\centering{}\includegraphics[scale=0.6]{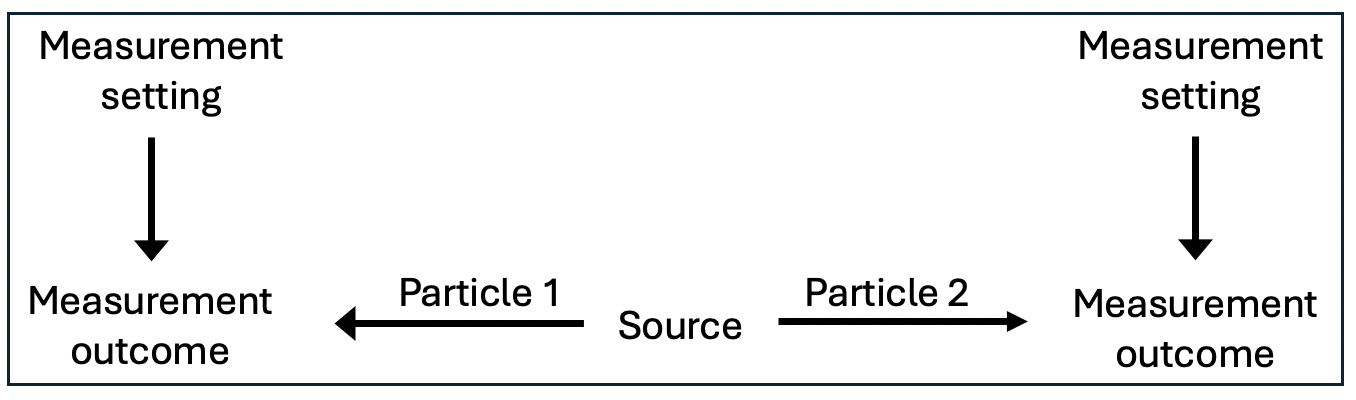}
\end{figure}
 Once they are very distant from each other, an observer, say Alice,
performs a measurement on the first particle, and another observer,
say Bob, performs a measurement on the second. Alice and Bob can each
choose between two distinct measurement settings. They can measure
either the position or the momentum of their respective particle.
So, Alice could choose to measure her particle's position while Bob
could choose to measure his particle's momentum, or vice versa, or
they could both make the same choice. Now, because the two particles
are maximally entangled, if Alice measures position, she can perfectly
predict the outcome of a position measurement performed by Bob; or,
if she measures momentum, she can perfectly predict the outcome of
a matching measurement on Bob\textquoteright s part too. The same
is true, symmetrically, for Bob. Alice\textquoteright s and Bob\textquoteright s
measurement outcomes will be perfectly correlated if they choose matching
measurement settings. Moreover, since the two particles are far apart
at the time of measurement, Alice\textquoteright s measurement should
not be able to affect any property of Bob\textquoteright s particle
instantaneously and vice versa. Clearly, if the two particles were
billiard balls in classical physics with perfectly correlated properties,
the situation would be completely unremarkable as each ball is taken
to possess position and momentum properties which can be read out
by measurement. If Alice were to measure the position or the momentum
of one of the two billiard balls, and she knew the other ball had
perfectly correlated properties, then of course she would be able
to predict the outcome of a matching measurement of the second ball.
Einstein, Podolsky, and Rosen assumed that perfect predictability
in the absence of any physical influence implies the existence of
some element of physical reality corresponding to the predictable
quantity. In the case of the maximally entangled particles, then,
the perfect predictability of either a position or a momentum measurement
of the distant particle, without any physical influence, suggests
that the distant particle possesses both position and momentum values,
just like perfectly correlated billiard balls.\footnote{And this is so, even though Alice would only be able to predict the
actual value of one property or the other, depending on which she
actually measures.} However, variables representing values for these unobserved measurements
are not present in the formalism of quantum mechanics. In fact, Heisenberg\textquoteright s
famous uncertainty principle, according to which one cannot measure
both a particle\textquoteright s position and its momentum with perfect
precision, is sometimes interpreted to imply that a particle does
not even \emph{possess} these properties unless or until they are
measured. Einstein, Podolsky, and Rosen concluded that quantum mechanics
must be incomplete. There must be some hidden facts, not yet acknowledged
by the standard theory, that determine what the various measurement
outcomes would be. 

Einstein, Podolsky, and Rosen's analysis leaves open the possibility
that the problem might be solved by postulating such hidden facts
(``hidden variables''). But a subsequent no-go result, Bell's theorem,
shows that this possibility is very limited (Bell 1964, Myrvold, Genovese,
and Shimony 2024). In particular, Bell's theorem shows that three
widely accepted physical theses are jointly inconsistent with the
predictions of quantum mechanics: 

\medskip{}

\noindent\textbf{Locality:} Measurement outcomes in one location
are probabilisticially independent of any measurement settings and
outcomes in spacelike-separated locations, conditional on any underlying
hidden variables.

\medskip{}

\noindent\textbf{Measurement independence:} Different observers'
measurement settings are probabilistically independent of each other
and of any property of the system that is being measured.\medskip{}

\noindent\textbf{Realism:} Any observer\textquoteright s measurement
outcomes correspond to objective facts about the system with well-defined
probabilities.

\medskip{}

In slightly more technical terms, Bell proved that if all three theses
are true of a given system, then the correlations between different
observers' measurement outcomes under different possible measurement
settings must satisfy certain constraints \textendash{} the so-called
Bell inequalities. But quantum mechanics predicts breaches of those
constraints. In essence, quantum-mechanical systems can display stronger
correlations than permitted if locality, measurement independence,
and realism are true. Therefore, at least one thesis must be false.

Let us briefly review the options. Physics has long emphasized locality
as a desirable assumption about physical reality, and locality holds
in Einstein's special and general theories of relativity. It is generally
considered undesirable to admit any correlations that cannot be accounted
for by local transmission mechanisms \textendash{} any ``spooky action
at a distance'', as Einstein famously put it. Still, some physicists
have interpreted Bell's theorem as showing that the physical world
is non-local. Maudlin (2014), for instance, argues that the ``proper
conclusion'' is that ``any world that displays violations of Bell\textquoteright s
inequality for experiments done far from one another must be non-local''
(p. 1). Indeed, several prominent interpretations of quantum mechanics
give up locality. The de Broglie-Bohm interpretation, for example,
depicts quantum-mechanical systems as deterministic systems whose
state evolves on the basis of their global particle configuration;
this can then account for non-local correlations (Goldstein 2025).
Similarly, Ney's wave function realism (Ney 2021) gives up locality,
by taking the state of a quantum system to be represented by a point
in a higher-dimensional configuration space, distinct from the conventional
three-dimensional physical space. From the perspective of our three-dimensional
space, we can thus see violations of locality. However, many physicists
consider violations of locality problematic. Newton endorsed the idea
of locality, even though he didn't use the term. He was famously unhappy
about the fact that in his theory gravitational forces have instantaneous
long-distance effects (Berkovitz 2016), a shortcoming that Einstein
eliminated in his general theory of relativity. In fact, Einstein
argued that giving up locality ``would make impossible the idea of
the existence of (quasi-)\linebreak{}
 closed systems and, thereby, the establishment of empirically testable
laws in the sense familiar to us'' (quoted and translated in Howard
1985, p. 188). 

This leads many to prefer a different response to Bell's theorem.
A second option is to give up measurement independence. If measurement
independence does not hold, Alice's and Bob's choices can be correlated
even though they are far apart. So-called superdeterministic interpretations
of quantum mechanics take this view (see, e.g., Hossenfelder and Palmer
2020). They assume that different observers' choices are correlated
due to some underlying hidden variable. The most radical version of
this view would say that their measurement choices are completely
predetermined. Measurement independence, however, is a deeply entrenched
assumption of science. Denying it has been thought to amount to a
denial that the experimenters have free will in choosing their measurement
settings, and many physicists worry that postulating the required
correlations between measurement settings would involve conspiratorial
thinking. Maudlin, for instance, argues that the postulated correlations
``would require a massive coincidence, on a scale that would undercut
the whole scientific method'' (2014, p. 22). Similarly, Sen and Valentini
(2020, p. 13) argue that ``at least as they are currently understood,
superdeterministic models are conspiratorial'', which leads them
``to conclude that superdeterminism, as a possible explanation of
the Bell correlations, is scientifically unattractive'' (ibid.).
A different way of relaxing measurement independence invokes the idea
of retrocausality (Price and Wharton 2015, Friederich and Evans 2023).
The proposal is roughly the following. While different observers'
measurement settings may be correlated with some underlying hidden
variable, the direction of causality doesn't go from the hidden variable
to the measurement settings but the other way around. So, it is the
measurement settings that causally affect the hidden variable, not
the hidden variable causally affecting the measurement settings. Postulating
a reversed direction of causality, however, is metaphysically costly
in its own way, and it is unclear that this move can fully address
the worry of conspiratorial thinking.\footnote{The worry is often presented as a ``fine-tuning worry'' (cf. Friederich
and Evans 2023). The postulated retrocausal effects must account for
the correlations predicted by quantum mechanics but without permitting
any superluminal signalling.}

The limitations of the foregoing responses to Bell's theorem and the
depth of the commitment to locality and measurement independence have
prompted some physicists to consider relaxing Bell's third thesis,
namely realism. Proponents of the so-called Copenhagen interpretation,
for example, fall into this camp. But this move faces some objections
too. Maudlin (2014) writes: ``it has become fashionable to say that
another way to avoid Bell\textquoteright s result and retain locality
is to abandon \emph{realism}. But such claims never manage to make
clear at the same time just what \textquoteleft realism\textquoteright{}
is supposed to be and just how Bell\textquoteright s derivation presupposes
it'' (p. 23). To deny realism, Maudlin suggests, we would have ``to
find any tacit physical claim {[}in Bell\textquoteright s argument{]}
that can be somehow denied'' (ibid.). As an example, he gives the
assumption ``that experiments have unique outcomes'', an assumption
denied by the Everett interpretation of quantum mechanics, but one
which, he notes, not only Bell but most people make ``when considering
the significance of Bell\textquoteright s result for the real world''
(ibid.). However, he warns, it would be unhelpful to deny realism
in a way that boils down to ``denying that the standard predictions
{[}of quantum mechanics{]} are actually correct'' (2014, p. 23).

But even if many physicists are reluctant to deny Bell's realism,
we should consider this response further, since the other responses
to Bell's theorem face serious objections. It is here that we will
re-encounter the challenge for objectivism. We will frame this challenge
with the help of a recent no-go result that extends Bell's theorem,
a ``heptalemma for quantum mechanics'', developed in DeBrota and
List (2026).

\section{The challenge for objectivism}

We have seen that if we wish to accommodate the predictions of quantum
mechanics while upholding locality and measurement independence, we
must reject Bell's thesis of realism. To explore what this means for
objectivism, we will make use of the insight (from DeBrota and List
2026) that Bell's realism thesis can be decomposed into several more
fine-grained subtheses. Specifically, it can be viewed as a conjunction
of four subtheses, each of which can be individually relaxed without
relaxing the others. The first expresses realism's basic ontic claim
and the subsequent three then impose further constraints on it. All
four subtheses are so deeply entrenched in mainstream science that
they are seldom explicitly articulated. Let us go through them. 

The basic ontic thesis expresses the core idea of realism, namely
that measurement outcomes are real in the sense that they correspond
to some kinds of facts (aspects of reality), perhaps in the weakest
imaginable way, to which one can assign probabilities. Unlike Bell's
original realism thesis, it imposes very few constraints on the nature
of those facts and especially does not build in any presumption of
objectivism. We use the wording from DeBrota and List (2026):

\medskip{}

\noindent\textbf{Measurement realism:} Any observer\textquoteright s
measurement outcomes correspond to facts over which there are well-defined
probabilities, where there is no presumption whether the facts are
objective or subjective, absolute or relative, coherent or incoherent.

\medskip{}
Note especially the ``no presumption'' clause. The subsequent three
theses then rule out that the facts in question could be relative,
incoherent, or subjective. We can think of these as capturing Bell's
objectivist assumptions. They match three of the theses introduced
in the earlier discussion of consciousness. We here paraphrase them
slightly.

\medskip{}

\noindent\textbf{Non-relationalism:} All facts, including those to
which measurement outcomes corespond, are of the absolute form ``such
and such is the case'', not of the relative form ``such and such
is the case, relative to such and such''.

\medskip{}

\noindent\textbf{Non-fragmentation: }The facts that hold at any given
world, as well as any assigments of probabilities, are jointly coherent.

\medskip{}

\noindent\textbf{One world: }Reality is exhausted by one objective
world, not multiple subjective worlds.

\medskip{}

Since these four subtheses jointly entail realism in Bell's original
sense, the Bell inequalities follow from their conjunction, plus locality
and measurement independence. To be precise, this presupposes non-solipsism,
since the EPR-style scenarios involved in the derivation of Bell's
theorem require more than one observer.

\medskip{}

\noindent\textbf{Non-solipsism:} There can be more than one observer.\medskip{}

We thus have the following result (DeBrota and List 2026):\medskip{}

\noindent\textbf{Heptalemma for quantum mechanics:} Locality, measurement
independence, measurement realism, non-relationalism, non-fragmentation,
one world, and non-solipsism are jointly inconsistent, given the predictions
of quantum mechanics. Any six of the seven theses are jointly consistent.

\medskip{}

For present purposes, we want to focus on one particular corollary
of this result. We obtain it if we group non-relationalism, non-fragmentation,
and one world together under the umbrella of objectivism and treat
locality and measurement independence as background assumptions. 

\medskip{}

\noindent\textbf{Corollary of the heptalemma:} Measurement realism,
non-solipsism, and objectivism are jointly inconsistent, given the
predictions of quantum mechanics and given the background assumptions
of locality and measurement independence. Any two of the initial three
theses \textendash{} measurement realism, non-solipsism, objectivism
\textendash{} are jointly consistent.

\medskip{}

Note that this restatement structurally mirrors the earlier no-go
result for consciousness. Although the present result is ultimately
a consequence of Bell's theorem, via the heptalemma, it makes Bell's
presupposition of non-solipsism explicit and exploits the fact that
Bell's realism thesis combines an ontic claim (measurement realism)
with an objectivist one (consisting of non-relationalism, non-fragmentation,
and one world).

Now, if we take measurement outcomes at face value \textendash{} treating
them as ontic and not merely epistemic, practical, or operational
\textendash{} then measurement realism is very reasonable. Giving
it up would entail a very anti-realist or purely operational understanding
of quantum mechanics, although some versions of the Copenhagen interpretation
appear to take that route (as discussed in DeBrota and List 2026).
Moreover, dropping non-solipsism is hardly plausible. Therefore, if
we are committed to locality and measurement independence, we must
reject objectivism. Putting this together, we get a remarkably similar
dialectic to the one in the case of consciousness, as illustrated
in Table 1.
\begin{table}[h]
\caption{No-go results for consciousness and quantum mechanics}

\centering{}%
\begin{tabular}{|c|c|c|}
\hline 
 & \textbf{Consciousness} & \textbf{Quantum mechanics}\tabularnewline
\hline 
\textbf{Ontic theses} & First-personal realism & Measurement realism\tabularnewline
\hline 
\multirow{3}{*}{\textbf{Objectivist theses}} & One world & One world\tabularnewline
\cline{2-3}
 & Non-fragmentation & Non-fragmentation\tabularnewline
\cline{2-3}
 & Non-relationalism & Non-relationalism\tabularnewline
\hline 
\textbf{Non-solipsist theses} & Non-solipsism & Non-solipsism\tabularnewline
\hline 
\multirow{2}{*}{\textbf{Background theses}} & \multirow{2}{*}{N/A} & Locality\tabularnewline
\cline{3-3}
 &  & Measurement independence\tabularnewline
\hline 
\end{tabular}
\end{table}

\section{Non-objectivist approaches to quantum mechanics}

Again, because objectivism is the conjunction of three theses, any
one of them can be relaxed on its own. Each of the three routes has
received some attention in recent work on quantum mechanics. For a
brief review, we will draw on the taxonomy developed in DeBrota and
List (2026), to which we refer the reader for more details. 

The first route upholds one world and non-fragmentation but gives
up non-relationalism. It assumes that the facts to which measurement
outcomes correspond are only relative rather than absolute. This is
the position taken by relational quantum mechanics (RQM), a prominent
approach pioneered by Rovelli (1996, 2025). In RQM, measurement outcomes
correspond to facts in a relative rather than absolute sense. Rovelli
(2025) notes that this involves a two-fold departure from classical
physics: ``(a) variables only take value at discrete interactions
and (b) the value a variable takes is only relative to the (other)
system affected by the interaction.'' So, in EPR-style scenarios,
measurement outcomes are not facts \emph{simpliciter}, but only facts
relative to Alice or facts relative to Bob. The total collection of
such relativized facts will then be perfectly coherent. RQM gives
a very general account of what the relativization parameter could
be: it need not be an observer in the conventional sense, but could
be any system. So, when we speak of ``facts relative to Alice'',
this means something like ``facts relative to Alice understood as
the system performing the measurement''. Generally, in RQM, facts
correspond to interactions between systems; every system-on-system
interaction can give rise to corresponding relative facts. Although
this is a departure from the objectivist worldview, it retains the
idea that there is a single world and that this world is coherent.
It just takes this world to be populated by relative rather than absolute
facts. 

The second non-objectivist route is the rejection of non-fragmentation.
This has also received some attention. Approaches that go in this
direction include quantum-logical approaches that offer a non-classical
account of the logical structure of possible events. According to
them, possible events do not form a Boolean algebra, where the operations
of conjunction (``and''), disjunction (``or''), and negation (``not'')
behave as in classical logic, but they form a lattice in which the
classical logical principle of distributivity fails (Wilce 2024).\footnote{Distributivity would require that ``$p$ and ($q$ or $r$)'' is
equivalent to ``($p$ and $q$) or ($p$ and $r$)'', and that ``$p$
or ($q$ and $r$)'' is equivalent to ``($p$ or $q$) and ($p$
or $r$)''.} Furthermore, some versions of QBism embrace the ``fragmentalist''
idea that quantum-mechanical reality might be only locally but not
globally coherent. Fuchs suggests the following analogy:
\begin{quote}
``An image that might be useful (but certainly flawed) comes from
Escher\textquoteright s various paintings of impossible objects. The
viewer would initially like to think of them as 2D projections of
a three-dimensional object; but he cannot. Now imagine how much worse
it would get if we were to have two viewers with two slightly different
paintings, each purporting to be a different perspective on `the'
impossible object. Since neither viewer can lift from his own 2D object
to a 3D one, there is no way to unify the pictures into a single whole.''
(Fuchs 2007, quoted in 2023, p. 127\textendash 128)
\end{quote}
Spelling out such a fragmentalist picture in a formally precise way
remains an ongoing project. Abramsky and Brandenburger (2011), for
instance, propose the use of mathematical tools from sheaf theory
to formalize a fragmentalist interpretation of quantum mechanics.

In some ways the most radical departure from objectivism is the rejection
of one world. This is similar to the route discussed above in relation
to consciousness, where the notion of a single objective world is
replaced by that of distinct subjective worlds: first-personally centred
worlds. An advantage of this, compared to the fragmentalist route,
is that it allows each of the subjective worlds to be internally consistent
and to be composed of facts in an absolute, non-relational sense.
The metaphysical cost is the proliferation of postulated worlds. 

At first, it may be tempting to think that the Everett interpretation
of quantum mechanics makes this move, as it is often called a ``many-worlds
interpretation'' and refers to a ``multiverse'' (Wallace 2012).
However, as noted in DeBrota and List (2026), this is not the best
way to understand that interpretation, since Everett still treats
the wave function as giving rise to a unified reality, which just
happens to be branching-tree-like. Chalmers describes this as a \textquotedblleft one-big-world
interpretation\textquotedblright{} and argues that it would be a misinterpretation
to attribute to Everett the view that ``the world literally splits
into many separate worlds every time a measurement is made'' (1996,
p. 347). An interpretation of quantum mechanics that better fits the
rejection of one world is the version of QBism described by Mermin
(2019). He writes: 
\begin{quote}
\textquotedblleft The fact is that \emph{my} science has a subject
(me) as well as an object (my world). \emph{Your} science has a subject
(you) as well as an object (your world). \emph{Alice}\textquoteright s
science has a subject (she) as well as an object (her world). I make
the same point three times to underline both the plurality of subjects,
and the plurality of worlds that each of us constructs on the basis
of our own individual experience.\textquotedblright{} (Mermin 2019,
p. 5)
\end{quote}
Alice's and Bob's records of the facts and probabilities in EPR-style
scenarios can then differ from one another without having to fit into
a single objective world. There can be distinct subjective worlds
and probability assignments for each of Alice and Bob, where these
are each internally coherent. In a similar spirit, Fuchs suggests
that ``{[}t{]}hat which surrounds each of us is more truly a pluriverse\textquotedblright{}
(2012, p. 11), and Pienaar (2024) argues that ``{[}a{]} single space-time
is too small for all of Wigner\textquoteright s friends'', referring
to Wigner's famous puzzle involving two observers. In DeBrota and
List (2026), we proposed the labels ``Fragmentalist QBism'' and
``Pluriverse QBism'' to distinguish a form of QBism that rejects
non-fragmentation from one that rejects one world. Both suggest significant
departures from the standard objectivist worldview.

\section{Comparing the cases of consciousness and quantum mechanics}

We have argued that consciousness and quantum mechanics both challenge
the classical objectivist worldview of science, and that they do so
in structurally similar ways. In the case of consciousness, if we
take first-personal facts at face value and find solipsism unpalatable,
then we must deny objectivism. In the case of quantum mechanics, if
we interpret measurement outcomes ontically rather than just epistemically,
practically, or operationally, and if we wish to uphold locality and
measurement independence, then we must also deny objectivism. In each
case, the denial can take three distinct forms: a relationalist form,
a fragmentalist form, or a many-subjective-worlds form. The first
is, in some respects, the least radical, as it retains the idea of
a single coherent world, while the last is, in some respects, the
most radical, as it abandons the idea of a single world altogether
and takes reality to be irreducibly plural. The middle option \textendash{}
fragmentalism \textendash{} retains the idea of a single world but
abandons the idea of global coherence. Whether this is metaphysically
attractive or unattractive is a matter for debate. 

Strictly speaking, there is no reason to require that the departure
from objectivism must take the same form in the case of consciousness
and in the case of quantum mechanics. One could in principle be a
fragmentalist about consciousness and a relationalist about quantum
mechanics, for instance. Nevertheless, considerations of theoretical
parsimony might speak in a favour of a structurally similar treatment
of the two cases. Let us now briefly compare the three possible ways
of aligning a non-objectivist theory of consciousness with a non-objectivist
interpretation of quantum mechanics. 

The first route is to embrace relationalism in both cases. In quantum
mechanics, as already noted, RQM offers a relational account of facts,
and it has some attraction as it retains the idea of a single coherent
world, albeit one populated by relational facts. In addition, the
research programme of RQM has been pursued for decades now and has
raised hopes of offering a viable way of relaxing Bell's realism without
overthrowing too much of the spirit of scientific realism. However,
with respect to consciousness, as already noted, giving a relational
account of first-personal facts would amount to a denial of first-personal
realism in the original sense. The fact that I, say Christian, am
having Christian's experiences is not merely the fact that relative
to Christian, it is true to say ``I am having Christian's experiences'',
but it is the fact that I am having Christian's experiences \emph{simpliciter}.
As noted, even the \emph{total} collection of relativized facts would
leave out the fact that \emph{I am having Christian's experiences
rather than anyone else's}, and so the relativized facts do not genuinely
capture the first-personal givenness of my experiences. Recall also
the difficulty of spelling out what exactly the relativization parameter
of such relationally re-interpreted first-personal facts is supposed
to be. If it is supposed to be a ``subject'', then we must still
clarify what a subject is and how subjects fit into reality. RQM seems
to offer a clearer solution to the problem of spelling out the relativization
parameter, because it implies that the relational facts of quantum
mechanics always correspond to relations between two systems.\footnote{However, Brukner (2021, p. 3) has raised some questions about this
aspect of RQM, arguing that ``not every quantum system can serve
as a measuring device or observer'' and that ``{[}w{]}hile every
measurement can be seen as an interaction between two systems ...,
not every interaction is a measurement''. } Arguably, then, the relationalist route works better in the case
of quantum mechanics than in the case of consciousness. But interestingly,
even in the case of quantum mechanics, some physicists, especially
in the QBist camp, have expressed misgivings about a relationalist
picture, in part because they find it too third-personal. Glick (2021),
taking a QBist perspective, notes that relational interpretations
``still aim to provide a description of external reality'' (p. 9),
albeit one where ``relational properties replace intrinsic (non-relational)
properties for physical systems'' (ibid.), and he points out that
``{[}w{]}hat QBism denies is that models in quantum theory should
be viewed as third-personal descriptions of external reality'' (p.
4), continuing: ``{[}f{]}or QBism, all of quantum theory is first-person
for the person who happens to be using it\textquotedblright{} (p.
9). 

The foregoing considerations might then lead us to set the relationalist
route aside and to prefer one of the other non-objectivist routes.
These are indeed favoured by most first-personal realists in the philosophical
debate on consciousness (recall the positions discussed by Fine 2005,
Merlo 2016, Lipman 2023, and one of us in List 2023, for example),
as well as by QBists in quantum mechanics (recall the quotes from
Fuchs 2007 and Mermin 2019, for example). The choice, then, is between
fragmentalism and a many-subjective-worlds theory. 

Fragmentalism retains the idea that first-personal facts and the facts
about measurement outcomes hold \emph{simpliciter} and not merely
in a relative sense, and it also retains the idea of a single world;
the usual identity ``reality = the world'' is saved. However, it
achieves this only at the cost of global incoherence. According to
fragmentalism, the collection of facts making up the world constitutes
an incoherent patchwork. There are locally coherent subcollections
of facts \textendash{} ``fragments'' \textendash{} but these do
not coherently fit together. Although tools from non-classical logic
might help us to make sense of this, the technical challenges are
still formidable, and fragmentalism would require us to abandon or
revise many of the entrenched formal tools we rely on in logic, science,
and philosophy: from possible worlds to probability functions and
probabilistic coherence. For this reason, a fragmentalist theory might
be rather costly overall.

A many-subjective-worlds theory, finally, retains the idea that any
``world'' is coherent, and that worlds are not merely constituted
by collections of relational facts. But it does postulate a proliferation
of worlds. To make sense of consciousness, we would have to postulate
as many first-personally-centred worlds as there are conscious subjects.
And to make sense of quantum mechanics, we would have to postulate
as many subjective worlds as there are observers, a point explicitly
affirmed by Mermin (2019) in the above-quoted passage: my world, your
world, Alice's word, and so on. 

The theoretical move from objective to subjective worlds may look
a bit less unpalatable, however, once we recognize that a many-subjective-worlds
approach still allows us to recover a useful notion of an objective
world. Objectivity can be defined as invariance under changes in the
subject or observer. If we specify a class of admissible such changes,
we can define as ``objective'' any fact that holds across those
changes. We can then introduce an equivalence relation on the set
of all subjective worlds by deeming any two subjective worlds to be
``third-personally equivalent'' or ``objectively equivalent''
if and only if they coincide with respect to all objective facts.
The equivalence classes with respect to this relation can be understood
as third-personal or impersonal worlds (List 2023). This way of thinking
about objectivity broadly echoes Husserl's idea that objectivity is
always rooted in intersubjectivity. Of course, the notion of a third-personal
or impersonal world becomes an abstraction under the present proposal,
and it is no longer fundamental, but from the perspective of the current
approach this is a feature rather than a bug. Worlds, according to
the present picture, are no longer fundamentally objective but fundamentally
subjective, centred around a subject or observer. According to the
many-subjective-worlds approach, this move is precisely what is needed
to do justice to the irreducible subjectivity of consciousness or
to the puzzles of quantum mechanics. Note that others, too, have proposed
a departure from the ``one world'' picture in order to make sense
of both the mind-body problem and quantum mechanics (see in particular
Vacariu 2005, 2016).

\section{From the book of the world to the library of reality}

To provide a more intuitive sense of the structure of each of the
different departures from objectivism we have discussed, it is helpful
to revisit the famous thought experiment of the ``book of the world''.\footnote{Here again, we draw on, and borrow some of the wording from, the exposition
in List (2025a). We also thank Philip Pettit for suggesting the ``library''
metaphor used below.} The ``book of the world'' is a gigantic book that contains a complete
and coherent record of all the facts that obtain in the world. It
exhaustively describes everything that is the case, up to the most
fine-grained detail, as it would be seen by a perfectly informed Olympian
observer. One might think of it as an omniscient being's full record
of what the world is like. A reader of the book could in principle
acquire a complete picture of reality. 

The objectivist worldview supports the idea of such a book. As Sider
(2011) summarizes the view, \textquotedblleft {[}t{]}here is an objectively
correct way to \textquoteleft write the book of the world\textquoteright\textquotedblright .
In practice, of course, it is very hard to write that book, but as
an ideal for science, the proponents of the objectivist worldview
say, the notion makes sense: the goal of science and philosophy is
to write that book. Of course, the objectivist worldview allows that
we may encounter all sorts of epistemological and practical difficulties
if we try to embark on this project. In reality, the book cannot be
written in finite time, and we will at most approximate it. Still,
objectivism implies that the ``book of the world'' is a coherent
ideal.

What we have seen is that consciousness and quantum mechanics each
challenge that ideal. Let's begin with consciousness. If we cannot
adequately accommodate consciousness in our worldview without postulating
irreducibly first-personal facts (and we don't think of these as being
merely relational), then there cannot exist a single objective book
of the world. As it is conventionally characterized, the book of the
world is written from a third-personal or impersonal perspective:
the ``view from nowhere''. Its narrator is an Olympian observer
describing the world from the outside. A book narrated like this would
fail to include any first-personal facts. It would thus be incomplete.
So, it would not be \emph{the book of the world}. For example, it
would describe Christian and his experiences in third-personal terms,
just as it would describe John's and everyone else\textquoteright s
experiences in third-personal terms, but it would leave out the first-personal
fact that I, say Christian, am having Christian\textquoteright s experiences
rather than anyone else\textquoteright s. A complete record of reality,
which includes subjective facts, could not be given in the form of
a unified objective book. Likewise, consider quantum mechanics. If
we accept locality and measurement independence as background assumptions
and we retain measurement realism as an ontic assumption, then it
will be impossible to provide a coherent and exhaustive enumeration
of all physical facts, non-relationally described. 

The three departures from objectivism we have discussed each correspond
to different ways of revising the ideal of ``the book of the world''.
The relational route corresponds to what is perhaps the least drastic
revision. According to relationalism, there could still be a kind
of ``book of the world'', but it would consist of relational facts.
So, it would not be a record of everything that is the case \emph{simpliciter},
but rather a record of what is the case relative to what. We might
call this ``the look-up table of the world'': a cross-tabulation
of which relational facts hold relative to which relativization parameters.
However, there would be a single such look-up table, and it would
be coherent. (We noted, however, that this may fail to accommodate
genuinely first-personal facts.)

The fragmentalist route would also retain a kind of book of the world,
but crucially it would no longer be unified and coherent. It would
no longer resemble the traditional genre of non-fiction or fiction,
familiar from textbooks or novels with a coherent story, but it would
be more akin to the genre of postmodern literature, where a book can
be explicitly fragmented and internally contradictory. A classic study
of postmodernist fiction describes the difference between modernist
and postmodernist fiction as follows: 
\begin{quote}
``Intractable epistemological uncertainty becomes at a certain point
ontological plurality or instability: push epistemological questions
far enough and they `tip over' into ontological questions.'' (McHale
1987, p. 11)
\end{quote}
According to fragmentalism, the book of the world may be compared
with a work of postmodernist fiction. Its internal fragmentation would
be a sign of reality's ``ontological plurality or instability''.
However, there is still a single book, albeit of a fragmented kind.

The many-subjective-worlds route, finally, would abandon the notion
of a single book of the world altogether. The notion of ``the book
of the world'' would be replaced by that of an entire library, consisting
of one subjective book for each subject or observer. We might call
it ``the library of reality''. Each book in that library would not
be written by an Olympian third-personal narrator, but it would reflect
the perspective of a particular subject or observer. The book of the
world written by one of us, which might be titled \textquotedblleft The
world as I found it\textquotedblright{} (as Ludwig Wittgenstein 1922
once proposed), would be different from that written by another. My
book, say Christian's, would be centred around my (Christian's) perspective,
while John's book would be centred around his; similarly for other
subjects. And in the case of our two experimenters performing quantum-mechanical
measurements, Alice's book would be different from Bob's. Of course,
different such books would still have some chapters in common, namely
those that cover only objective facts \textendash{} facts that are
invariant under shifts in perspective. But the different books would
come apart in their subjective or observer-specific content, which
could be substantial. They could not be coherently merged into a single
book because such a merged book could not be narrated from a unified
perspective. At best, such a hypothetical merger would take us back
to the genre of postmodernist literature.

In short, while objectivism identifies ``writing the book of the
world'' (Sider 2011) as an ideal for science and philosophy, the
three departures from objectivism revise that ideal: they replace
``the book of the world'' with a look-up table, a piece of postmodernist
fiction, or an entire library, respectively. We may or may not like
these alternative ideals, and we may find them less parsimonious than
the ideal of a single coherent book, but if we take the challenges
from consciousness and quantum mechanics seriously, we may have to
come to terms with one of these alternatives.

\section{Concluding remarks}

We have compared the ways consciousness and quantum mechanics each
challenge the classical objectivist worldview of science. At least
under certain assumptions, they each conflict with a package of metaphysical
theses that jointly make up that worldview. There is very little room
for avoiding the conflict. If we set aside the option of embracing
solipsism, in the case of consciousness, the only way to avoid the
conflict would be to deny first-personal realism altogether. The explanatory
burden would then be to show how this can be done without lapsing
into some form of illusionism. And in the case of quantum mechanics,
the only way to avoid the conflict would be to reject either locality
or measurement independence or measurement realism, and thereby to
go against one of three core assumptions to which many scientists
are deeply committed.

Consequently, addressing the challenge may require a departure from
objectivism. We have mapped out the three main possibilities: the
relationalist route, the fragmentalist route, and the many-subjective-worlds
route. Although we have not definitively argued for one of these routes,
one of us has a preference for the last route (along the lines explained
in List 2023). This prompts the methodological question, recognized
by Strawson (1966), of what might count as evidence against the conventional
thesis of one objective world. If, as Moore (2016, p. 936) writes,
``the principal issue'' is ``whether \textquoteleft reality\textquoteright{}
is unified, or whether, on the contrary, it consists of more than
one \textquoteleft world\textquoteright \,'', one might wonder what
could speak in favour of the many-worlds option. It is unlikely that
empirical evidence alone could adjudicate the issue. Rather, the answer,
we think, must ultimately come from an inference to the best explanation.
The key consideration is which metaphysical worldview works best overall,
given the explanatory challenges we are faced with. If accounting
for the \emph{explananda} of consciousness and quantum mechanics turns
out to conflict with objectivism, and we are unwilling to give up
any of the other assumptions leading to that conflict (as detailed
in Table 1), then we must relax either non-relationalism, or non-fragmentation,
or one world. Which of these routes we should take comes down to the
question of which is explanatorily best. It might be that one of the
three non-objectivist routes is ultimately more elegant, more illuminating,
or better able to accommodate a broad range of phenomena than the
others. That would then speak for the route in question. 

In conclusion, it is worth briefly returning to phenomenology. The
phenomenologists anticipated many of the issues discussed here (again,
see French 2023). Husserl wrote: 
\begin{quote}
``A purely Objective science aims at a theoretical cognizing of Objects,
not in respect of such subjectively relative determinations as can
be drawn from direct sensuous experience, but rather in respect of
strictly and purely Objective determinations: determinations that
obtain for everyone and at all times, or in respect of which according
to a method that everyone can use, there arise theoretical truths
having the character of \textquoteleft truths in themselves \textendash{}
in contrast to mere subjectively relative truths\textquoteright .''
(1969, p. 38, quoted in French 2023, p. 168)
\end{quote}
And he famously criticized that picture of science, arguing that it
is deeply limited. It focuses only on third-personally quantifiable
abstractions and loses sight of the world of experience of the observer
or subject, their ``lifeworld'', which he thinks should be central
to any enquiry. For structural reasons, in particular, conscious experience
cannot be adequately accommodated by an objectivist approach. Husserl's
critique of the objectivist physicalism of science culminates in his
unfinished work \emph{The Crisis of European Sciences and Transcendental
Phenomenology} (Husserl 1970). London and Bauer (1939) took inspiration
from Husserl's ideas and extended the phenomenological critique to
quantum mechanics, as discussed by French (2023). London thought that
quantum mechanics requires us to abandon the quest for an objective
description of reality and to come to terms with the centrality of
the subject:
\begin{quote}
``Most of us today feel that this necessary abandonment of a purely
objective description of Nature is a profound change in the physical
concept of the world. We feel it as a painful limitation of our right
to truth and clarity, that our symbols and formulas and the pictures
connected with them do not represent an object independent of the
observer but only the relation of subject to object. But is this relation
not basically the one true reality that we know?'' (London, quoted
in French 2023, p. 99)
\end{quote}
French (2023) goes so far as to suggest that quantum mechanics ``re-unites
nature and consciousness and London and Bauer\textquoteright s \textquoteleft little
book\textquoteright{} {[}on the interpretation of quantum mechanics{]}
completes Husserl\textquoteright s project'' (p. 185) (cf. Pienaar
2025). Be that as it may, we think the no-go results we have discussed
here reinforce the phenomenologists' challenge for the objectivist
worldview. Even if the problems of consciousness and quantum mechanics
are distinct, and we do not simply claim to reduce one mystery to
the other, they raise some structurally similar challenges. Comparing
those challenges may yield useful lessons on how one might pursue
science without objectivism.

\section*{References}
\begin{lyxlist}{XXX}
\item [{Abramsky,}] S., and A. Brandenburger (2011). ``The sheaf-theoretic
structure of non-locality and contextuality.'' \emph{New Journal
of Physics} 13: 113036.
\item [{Baker,}] L. R. (2013) \emph{Naturalism and the First-Person Perspective}.
Oxford: Oxford University Press.
\item [{Bell,}] J.S. (1964) ``On the Einstein Podolsky Rosen paradox.''
\emph{Physics Physique Fizika} 1: 195\textendash 200.
\item [{Berghofer,}] P., and H. A. Wiltsche (eds.) (2023) \emph{Phenomenology
and QBism: New approaches to quantum mechanics}. New York: Routledge.
\item [{Berkovitz,}] J. (2016) ``Action at a Distance in Quantum Mechanics.''
In E. N. Zalta (ed.), \emph{The Stanford Encyclopedia of Philosophy}
(Spring 2016 Edition). \textless https://plato.stanford.edu/archives/spr2016/entries/qm-action-distance/\textgreater .
\item [{Brukner,}] C. (2021) ``Qubits are not observers \textendash{}
a no-go theorem.'' arXiv:2107.03513v1.
\item [{Builes,}] D. (2024) ``Eight Arguments for First-Person Realism.''
\emph{Philosophy Compass} 19: e12959.
\item [{Chalmers,}] D. (1995) ``Facing up to the problem of consciousness.''
\emph{Journal of Consciousness Studies} 2: 200\textendash 219. 
\item [{Chalmers,}] D. (1996) \emph{The Conscious Mind}. New York: Oxford
University Press.
\item [{Chalmers,}] D., and K. J. McQueen (2022) ``Consciousness and the
Collapse of the Wave Function.'' In S. Gao (ed.), \emph{Consciousness
and Quantum Mechanics}, pp. 11\textendash 63. New York: Oxford University
Press.
\item [{Conitzer,}] V. (2020) ``The Personalized A-Theory of Time and
Perspective.'' \emph{Dialectica} 74(1): 3\textendash 31.
\item [{DeBrota,}] J., and C. List (2026) ``A heptalemma for quantum mechanics.''
\emph{Foundations of Physics} 56: 24. 
\item [{Declos,}] A. (2019) ``Goodman\textquoteright s Many Worlds.''
\emph{Journal for the History of Analytical Philosophy} 7: 6.
\item [{Einstein,}] A., B. Podolsky, B., and N. Rosen (1935) ``Can quantum-mechanical
description of physical reality be considered complete?'' \emph{Physical
Review} 47(10): 777\textendash 780.
\item [{Fine,}] K. (2005) ``Tense and Reality.'' In Modality and Tense:
\emph{Philosophical Papers}, pp. 261\textendash 320. Oxford: Oxford
University Press.
\item [{Frankish,}] K. (2016) ``Illusionism as a Theory of Consciousness.''
\emph{Journal of Consciousness Studies} 23: 11\textendash 39.
\item [{French,}] S. (2023) \emph{A Phenomenological Approach to Quantum
Mechanics: Cutting the Chain of Correlations}. Oxford: Oxford University
Press.
\item [{Friederich,}] S., and P. W. Evans (2023) ``Retrocausality in quantum
mechanics.'' In E. N. Zalta and U. Nodelman (eds.), \emph{The Stanford
Encyclopedia of Philosophy} (Winter 2023 Edition). \textless https://plato.stanford
.edu/archives/win2023/entries/qm-retrocausality/\textgreater .
\item [{Fuchs,}] C. A. (2007) ``Delirium quantum: Or, where I will take
quantum mechanics if it will let me.'' In G. Adenier, C. A. Fuchs,
A. Yu. Khrennikov (eds.), \emph{Foundations of probability and physics
\textendash{} 4, AIP Conference Proceedings}, Vol. 889, pp. 438\textendash 462.
Melville, NY: American Institute of Physics. 
\item [{Fuchs,}] C. A. (2012) ``Interview with a Quantum Bayesian.''
arXiv:1207.2141.
\item [{Fuchs,}] C. A. (2023) ``QBism, where next?'' In P. Berghofer
and H. A. Wiltsche (eds.), \emph{Phenomenology and QBism: New approaches
to quantum mechanics}, pp. 78\textendash 143. New York: Routledge.
\item [{Gabriel,}] M. (2015) \emph{Why the World Does Not Exist}. Cambridge:
Polity.
\item [{Gao,}] S. (ed.) (2022) \emph{Consciousness and Quantum Mechanics}.
New York: Oxford University Press.
\item [{Glick,}] D. (2021) ``QBism and the limits of scientific realism.''
\emph{European Journal for Philosophy of Science} 11: 53.
\item [{Goldstein,}] S. (2025) ``Bohmian Mechanics.'' In E. N. Zalta
and U. Nodelman (eds.), \emph{The Stanford Encyclopedia of Philosophy}
(Fall 2025 Edition). \textless https://plato.stanford.edu/archives/fall2025/entries/qm-bohm/\textgreater .
\item [{Goodman,}] N. (1978) \emph{Ways of Worldmaking}. Indianapolis:
Hackett.
\item [{Goodman,}] N. (1984) \emph{Of Mind and Other Matters}. Cambridge,
MA: Harvard University Press.
\item [{Hameroff,}] S., and R. Penrose (2014) ``Consciousness in the universe:
A review of the \textquoteleft Orch OR\textquoteright{} theory.''
\emph{Physics of Life Reviews} 11(1): 39\textendash 78.
\item [{Hare,}] C. (2007) ``Self-bias, Time-bias, and the Metaphysics
of Self and Time.'' \emph{Journal of Philosophy} 104(7): 350\textendash 373.
\item [{Hellie,}] B. (2013) ``Against Egalitarianism.'' \emph{Analysis}
73(2): 304\textendash 320.
\item [{Honderich,}] T. (2014) \emph{Actual Consciousness}. Oxford: Oxford
University Press.
\item [{Hossenfelder,}] S., and T. Palmer (2020) ``Rethinking Superdeterminism.''
\emph{Frontiers in Physics: Statistical and Computational Physics}
8: 139.
\item [{Howard,}] D. (1985) ``Einstein on locality and separability.''
Studies in History and Philosophy of Science Part A 16(3): 171\textendash 201.
\item [{Husserl,}] E. (1969) \emph{Formal and Transcendental Logic}. Transl.
by D. Cairns. The Hague: Nijhoff.
\item [{Husserl,}] E. (1970) \emph{The Crisis of European Sciences and
Transcendental Phenomenology}. Transl. by D. Carr. Evanston: Northwestern
University Press.
\item [{Kant,}] I., transl. by Smith, N. K. (1929) \emph{Immanuel Kant's
Critique of Pure Reason}. London (Macmillan).
\item [{Kirk,}] R. (2023) ``Zombies.'' In E. N. Zalta and U. Nodelman
(eds.), \emph{The Stanford Encyclopedia of Philosophy} (Fall 2023
Edition). \textless https://plato.stanford.edu/archives/fall2023/entries/zombies/\textgreater .
\item [{Lewis,}] D. (1986) \emph{On the Plurality of Worlds}. Oxford: Blackwell.
\item [{Lipman,}] M. (2023) ``Subjective Facts about Consciousness.''
\emph{Ergo: An Open Access Journal of Philosophy} 10: 530\textendash 553.
\item [{List,}] C. (2023) ``The Many-Worlds Theory of Consciousness.''
\emph{Noûs} 57: 316\textendash 340.
\item [{List,}] C. (2025) ``A quadrilemma for theories of consciousness.''
\emph{The Philosophical Quarterly} 75(3): 1026\textendash 1048.
\item [{List,}] C. (2025a) ``Consciousness reveals there's no single objective
world: Science as we know it won\textquoteright t explain consciousness.''
iai.tv. \textless https://iai.tv/articles/consciousness-reveals-reality-cannot-be-described-auid-3151/\textgreater . 
\item [{London,}] F., and E. Bauer (1939) \emph{La théorie de l\textquoteright observation
en mécanique quantique}. Paris: Hermann.
\item [{Maudlin,}] T. (2014) ``What Bell Did.'' \emph{Journal of Physics
A: Mathematical and Theoretical} 47(42): 424010.
\item [{McHale,}] B. (1987) \emph{Postmodernist Fiction}. London: Routledge.
\item [{Merlo,}] G. (2016) ``Subjectivism and the Mental.'' \emph{Dialectica}
70(3): 311\textendash 342.
\item [{Mermin,}] N. D. (2019) \textquotedblleft Making better sense of
quantum mechanics.\textquotedblright{} \emph{Reports on Progress in
Physics} 82: 012002.
\item [{Moore,}] A. W. (2016) ``One World.'' European Journal of Philosophy
24(4): 934\textendash 945.
\item [{Mørch,}] H. H. (2024) \emph{Non-Physicalist Theories of Consciousness}.
Cambridge: Cambridge University Press.
\item [{Myrvold,}] W., M. Genovese, and A. Shimony (2024) ``Bell\textquoteright s
Theorem.'' In E. N. Zalta and U. Nodelman (eds.), \emph{The Stanford
Encyclopedia of Philosophy} (Spring 2024 Edition). \textless https://plato.stanford.edu/archives/spr2024/entries/bell-theorem/\textgreater .
\item [{Nagel,}] T. (1965) ``Physicalism.'' \emph{The Philosophical Review}
74: 339\textendash 356.
\item [{Nagel,}] T. (1974) ``What Is It Like to Be a Bat?'' \emph{The
Philosophical Review} 83: 435\textendash 450.
\item [{Nagel,}] T. (1986) \emph{The View From Nowhere}. New York: Oxford
University Press.
\item [{Ney,}] A. (2021) \emph{The world in the wavefunction: A metaphysics
for quantum physics}. Oxford (Oxford University Press).
\item [{Penrose,}] R. (1994) \emph{Shadows of the Mind: A Search for the
Missing Science of Consciousness}. Oxford: Oxford University Press.
\item [{Pienaar,}] J. L. (2024) ``A single space-time is too small for
all of Wigner's friends.'' arXiv:2312.11759.
\item [{Pienaar,}] J. L. (2025) ``French on London and Bauer, and QBism.''
\emph{Studies in History and Philosophy of Science} 114: 102084.
\item [{Price,}] H., and K. Wharton (2015). ``Disentangling the quantum
world.'' \emph{Entropy} 17(11): 7752\textendash 7767.
\item [{Ratcliffe,}] M. (2002). ``Husserl and Nagel on Subjectivity and
the Limits of Physical Objectivity.'' \emph{Continental Philosophy
Review} 35: 353\textendash 377.
\item [{Rovelli,}] C. (1996) ``Relational quantum mechanics.'' \emph{International
Journal of Theoretical Physics} 35(8): 1637\textendash 1678.
\item [{Rovelli,}] C. (2025) ``Relational quantum mechanics.'' In E.
N. Zalta and U. Nodelman (eds.) \emph{The Stanford Encyclopedia of
Philosophy} (Spring 2025 Edition). \textless https://plato.stanford.edu/archives/spr2025/entries/qm-relational/\textgreater .
\item [{Sen,}] I., and A. Valentini (2020) ``Superdeterministic hidden-variables
models II: conspiracy.'' \emph{Proceedings of the Royal Society A}
476 (2243): 20200214.
\item [{Sider,}] T. (2011) \emph{Writing the Book of the World}. Oxford:
Oxford University Press.
\item [{Strawson,}] P. F. (1966) The Bounds of Sense. London: Methuen.
\item [{Vacariu,}] G. (2005) ``Mind, Brain, and Epistemologically Different
Worlds.'' \emph{Synthese} 147(3): 515\textendash 548. 
\item [{Vacariu,}] G. (2016) \emph{Illusions of Human Thinking: On Concepts
of Mind, Reality, and Universe in Psychology, Neuroscience, and Physics}.
Wiesbaden: Springer.
\item [{Wallace,}] D. (2012) ``The emergent multiverse: Quantum theory
according to the Everett interpretation.'' Oxford: Oxford University
Press.
\item [{Wilce,}] A. (2024) ``Quantum Logic and Probability Theory.''
In E. N. Zalta and U. Nodelman (eds.), \emph{The Stanford Encyclopedia
of Philosophy} (Summer 2024 Edition). \textless https://plato.stanford.edu/archives/sum2024/entries/qt-quantlog/\textgreater .
\item [{Wiltsche,}] H. A., and P. Berghofer (eds.) (2020) \emph{Phenomenological
Approaches to Physics}. Heidelberg: Springer.
\item [{Wittgenstein,}] L. (1922) \emph{Tractatus Logico-Philosophicus.}
Transl. by C. K. Ogden. London: Routledge. 
\item [{Zahavi,}] D. (2025) ``Edmund Husserl.'' In E. N. Zalta and U.
Nodelman (eds.), \emph{The Stanford Encyclopedia of Philosophy} (Winter
2025 Edition). \textless https://plato.stanford.edu/archives/win2025/entries/husserl/\textgreater .
\end{lyxlist}

\end{document}